\documentclass{aa}
\usepackage{epsfig,graphics}

\begin{document}

\title{Magnetic field measurements of O stars with FORS\,1 at the VLT\thanks{Based on observations obtained at the European Southern Observatory,
Paranal, Chile (ESO programmes 075.D-0432(A), 078.D-0330(A), 079.D-0241(A), 080.D-0383(A)).}}

\author{
S. Hubrig\inst{1}
\and
M. Sch\"oller\inst{1}
\and
R.S. Schnerr\inst{2,3}
\and
J.~F. Gonz\'alez\inst{4}
\and
R. Ignace\inst{5}
\and
H.~F. Henrichs\inst{2}
}

\institute{
European Southern Observatory, Casilla 19001, Santiago 19, Chile
\and
Astronomical Institute ``Anton Pannekoek'', University of Amsterdam, Kruislaan 403, 1098 SJ Amsterdam, The Netherlands
\and
SRON, Netherlands Institute for Space Research, Sorbonnelaan 2, 3584 CA Utrecht, The Netherlands 
\and
Complejo Astron\'omico El Leoncito, Casilla 467, 5400 San Juan, Argentina
\and
Department of Physics, Astronomy, \& Geology, East Tennessee State University, Johnson City, TN
 37614, USA
}

\date{Received date / Accepted date}

\abstract
{The presence of magnetic fields in O-type stars has been suspected for a long time.
The discovery of such fields would explain a wide range of well documented enigmatic phenomena in massive stars,
in particular cyclical wind variability, H$\alpha$ emission variations, chemical peculiarity, narrow X-ray emission lines and 
non-thermal radio/X-ray emission.}
{To investigate the incidence of magnetic fields in O stars,
we acquired 38 new spectropolarimetric observations with FORS\,1 (FOcal Reducer low dispersion Spectrograph)
mounted on the 8-m Kueyen telescope of the VLT.
}
{
Spectropolarimetric observations have been obtained at different phases for a sample of 13 O stars.
10 stars were observed in the spectral range 348--589\,nm,
HD\,36879 and HD\,148937 were observed in the spectral region 325--621\,nm,
and HD\,155806 was observed in both settings.
To prove the feasibility of the FORS\,1 spectropolarimetric mode for the measurements 
of magnetic fields in hot stars, we present in addition 12 FORS\,1 observations 
of the mean longitudinal magnetic field in $\theta^1$\,Ori\,C and compare them with measurements 
obtained with the MuSiCoS, ESPaDOnS and Narval spectropolarimeters.
}
{
Most stars in our sample which have been observed on different nights show a change of the magnetic field polarity, 
but a field at a significance level of 3$\sigma$ has been detected only in four stars, HD\,36879, HD\,148937, HD\,152408, 
and HD\,164794. The largest longitudinal magnetic field, $\langle$$B_z$$\rangle$\,=\,$-$276$\pm$88\,G, was 
detected in the Of?p star HD\,148937.
We conclude that large-scale organised magnetic fields with polar field strengths larger than 1\,kG are 
not widespread among O-type stars.
}
{}


\titlerunning{Magnetic field measurements of O stars with FORS\,1/VLT}
\authorrunning{S.\ Hubrig et al.}
\maketitle



\section{Introduction}
\label{sect:intro}
Massive stars usually end their evolution with a final supernova explosion, producing neutron stars 
or black holes. The initial masses of these stars range from $\sim$9--10\,M$_\odot$ to 
100\,M$_\odot$ or more, which correspond to spectral types earlier than about B2.
Magnetic O stars with masses larger than 30\,M$_\odot$  and their WR descendants
have been suggested as progenitors of magnetars (Gaensler et al.\ \cite{gaensler05}).
Contrary to the case of Sun-like stars, the
magnetic fields of stars on the upper main sequence (Ap/Bp stars),
white dwarfs, and neutron stars are dominated by large spatial
scales and do not change on yearly time scales. In each of these
classes there is a wide distribution of magnetic field strengths,
but the distribution of magnetic fluxes appears to be similar in
each class, with maxima of $\Phi_\mathrm{max}=\pi R^2B\sim
10^{27-28}\mathrm{G~cm^2}$ (Reisenegger \cite{reisenegger01}, Ferrario \& Wickramasinghe \cite{ferrario05}),
arguing for a fossil field
whose flux is conserved along the path of stellar evolution.
Braithwaite \& Spruit (\cite{braithwaite04}) confirmed through
simulations that there are stable MHD configurations that might
account for long-lived, ordered fields in these types of stars.

However, very little is known about the existence, origin, and role of magnetic fields
in massive O and Wolf-Rayet stars.
The lack of information
is especially disturbing because magnetic fields may have
paramount influence on the stellar evolution of high-mass stars. 
Maeder \& Meynet (\cite{maeder05}) examined the effect of  magnetic fields on the
transport of angular momentum and chemical mixing, and they found that the
potential influence on the  evolution of massive stars is dramatic.

Indirect observational evidence for the presence of magnetic fields are the many unexplained phenomena observed in massive stars,
that are thought to be related to magnetic fields.
One of the main indications that massive stars have magnetic fields is the cyclic behaviour on a rotational timescale observed
in the UV wind lines (e.g.\ Henrichs et al.\ \cite{henrichs:2005}).
Other indications are variability
observed in the H and He lines
(Moffat \& Michaud \cite{moffat:1981}, Stahl et al.\ \cite{stahl:1996}, Rauw et al.\ \cite{rauw:2001}),
narrow X-ray emission lines (Cohen et al.\ \cite{cohen:2003}, Gagn\'e et al.\ \cite{gagne:2005b})
and the presence of non-thermal radio emission (Bieging et al.\ \cite{bieging:1989}, Scuderi et al.\ \cite{scuderi:1998},
Schnerr et al.\ \cite{schnerr:2006c}).

Direct measurements of the magnetic field strength in massive stars using spectropolarimetry to determine the Zeeman splitting
of the spectral lines is difficult, as only fewer spectral lines are available for the measurements and which are usually strongly broadened 
by rapid rotation.
So far a magnetic field has only been found in three O stars, $\theta^1$\,Ori\,C, HD\,155806 and HD\,191612
(Donati et al.\ \cite{donati:2002},  Hubrig et al.\ \cite{hubrig07}, Donati et al.\ \cite{donati:2006})
and in a handful of early B-type stars
(Henrichs et al.\ \cite{henrichs:2000a}, Neiner et al.\ \cite{coralie:2003a}, Neiner et al.\ \cite{coralie:2003b},
Neiner et al.\ \cite{coralie:2003c}, Hubrig et al.\ \cite{hubrig:2006}, Donati et al.\ \cite{donati:2006b}, Hubrig et al.\ 2007).
In this paper we present the results of the measurements of magnetic fields in 13 O type stars using FORS\,1 at the VLT 
in spectropolarimetric mode.
Our observations and the data reduction are described in Sect.~\ref{sect:observations}, the obtained results in 
Sect.~\ref{sect:results} and their discussion is presented in Sect.~\ref{sect:discussion}.

\section{Observations and data reduction}
\label{sect:observations}

\begin{table}
\caption{
Target stars discussed in this paper.
Spectral types are from Ma{\'{\i}}z-Apell{\'a}niz et al.\ (\cite{maiz:2004}), $v\sin i$ values
are taken from Howarth et al.\ (\cite{howarth97}).
For two stars, HD\,135240 and HD\,167771, 
not considered by Howarth et al.\ (\cite{howarth97}), the $v\sin i$ values are from the Bright 
Star Catalogue (Hoff\/leit \& Jaschek \cite{bsc5:1991}).
The $v\sin i$ value for HD\,148937 was recently reported by  Naz\'e et al.\ (\cite{naze08})
}
\label{tab:targetlist}
\centering
\begin{tabular}{rlclc}
\hline
\hline
HD     & Other    & V& Spectral & $v\sin i$\\
number & name     &   &  type    & [km s$^{-1}$]\\
\hline
36879  & BD$+$21 899 & 7.6 &  O7 V(n)& 163\\
112244 & HR 4908 &5.3  & O8.5 Iab(f)      & 147\\
135240 & $\delta$\,Cir &5.1 & O7.5 III((f))    & 189\\
135591 & HR 5680 & 5.3& O7.5 III((f))    & 78\\
148937 & CD$-$47 10855 &6.8 & O6.5 f?p         & 45\\
151804 &HR 6245	&5.2 & O8 Iaf           & 104\\
152408 & HR 6272 & 5.9 & O8: Iafpe        & 85\\
155806 &HR 6397	&5.6 & O7.5 V[n]e       & 91\\
162978 &63 Oph 	& 6.2 & O7.5 II((f))     &  86\\
164794 & 9\,Sgr  & 5.9& O4 V((f))        & 70\\
167263 & 16\,Sgr &6.0  & O9.5 II-III((n)) & 99\\
167771 & HR 6841 &6.5 & O7 III:(n)((f)) &  90\\
188001 & 9\,Sge  &6.2 & O7.5 Iaf         & 93\\
\hline
\end{tabular}
\end{table}

The major part of the observations reported here have been carried out
between March and August of 2005 in service mode
at the European Southern Observatory with FORS\,1 
mounted on the 8-m Kueyen telescope of the VLT. This multi-mode instrument is equipped with
polarisation analyzing optics comprising super-achromatic half-wave and quarter-wave 
phase retarder plates, and a Wollaston prism with a beam divergence of 22$\arcsec$  in 
standard resolution mode. 11 O-type stars have been observed in 2005
with the GRISM 600B in the wavelength range 3480--5890\,\AA{}
to cover all hydrogen Balmer lines from H$\beta$ to the Balmer jump. 
Their selection was based on the extensive study of wind variability in O and B stars
using the IUE data archive by ten Kulve (\cite{Kulve04}), anomalous X-ray behaviour and brightness.
The spectral types of the studied stars are listed in Table~\ref{tab:targetlist} and the observed FORS\,1 
spectra in integral light are presented in Fig.~\ref{fig:spectra}.
The observation  of HD\,36879 has been obtained at the beginning of September 2007 and two more 
observations, one for the star HD\,148937 and another one for the star HD\,155806,
have been obtained at the end of March 2008.
These observations have been carried out  with the GRISM 600B and a new mosaic detector
with blue optimised E2V chips, which was implemented in FORS\,1 at the beginning of April 2007.
It has a  pixel size of 15\,$\mu$m (compared to 24\,$\mu$m for the
previous Tektronix chip) and higher efficiency
in the wavelength range below 6000\,\AA{}.
With the new mosaic detector
and the grism 600B we are also able now to cover a much larger
spectral range, from 3250 to 6215\,\AA{}.

12 observations of the magnetic O star $\theta^1$\,Ori\,C, distributed over 
the rotational period, have been obtained in 2006 with GRISM 600R in the 
wavelength range 5240--7380\,\AA{}. 
In all observations the narrowest slit width of 0$\farcs$4 was used to obtain a spectral resolving power 
of $R\sim2000$ with GRISM 600B and $R\sim3000$ with GRISM 600R.

The mean longitudinal magnetic field, $\langle B_z\rangle$, has been derived using 

\begin{equation}
\frac{V}{I} = -\frac{g_{\rm eff} e \lambda^2}{4\pi{}m_ec^2}\ \frac{1}{I}\ \frac{{\rm d}I}{{\rm d}\lambda} \left<B_z\right>,
\label{eqn:one}
\end{equation}

where $V$ is the Stokes parameter which measures the circular polarisation,
$I$ is the intensity in the unpolarized spectrum,
$g_{\rm eff}$ is the effective Land\'e factor,
$e$ is the electron charge,
$\lambda$  is the wavelength expressed in \AA{},
$m_e$ the electron mass,
$c$ the speed of light, ${{\rm d}I/{\rm d}\lambda}$ is the derivative of 
Stokes $I$,
and $\left<B_z\right>$ is the mean longitudinal field expressed in Gauss. 
To minimize the cross-talk effect we executed the sequence
+45$-$45, +45$-$45, +45$-$45 etc. and
calculated the values $V/I$ using:

\begin{equation}
\frac{V}{I} =
\frac{1}{2} \left\{ \left( \frac{f^{\rm o} - f^{\rm e}}{f^{\rm o} + f^{\rm e}} \right)_{\alpha=-45^{\circ}}
- \left( \frac{f^{\rm o} - f^{\rm e}}{f^{\rm o} + f^{\rm e}} \right)_{\alpha=+45^{\circ}} \right\},
\label{eqn:two}  
\end{equation}

where $\alpha$ gives the position angle of the retarder waveplate
and $f^{\rm o}$ and $f^{\rm e}$ are ordinary and
extraordinary beams, respectively. Stokes $I$ values have been obtained from the sum of the 
ordinary and extraordinary beams.
To derive $\left<B_z \right>$, a least-squares technique has been
used to minimize the expression

\begin{equation}
\chi^2 = \sum_i \frac{(y_i - \left<B_z \right> x_i - b)^2}{\sigma_i^2}
\label{eqn:three}  
\end{equation}

where, for each spectral point $i$, $y_i = (V/I)_i$,
$x_i = -\frac{g_{\rm eff} e \lambda_i^2}{4\pi{}m_ec^2}\ (1/I\ \times\ {\rm d}I/{\rm d}\lambda)_i$
and $b$ is a constant term that, assuming that Eq.~\ref{eqn:one} is
correct, approximates the fraction of instrumental polarisation not removed
after the application of Eq.~\ref{eqn:two} to the observations.
During the commissioning of FORS\,1, this 
instrumental polarisation term was found to be wavelength independent.
A wavelength dependent instrumental polarisation  would also be visible 
in the $V$/$I$  spectra, but we do not see anything like this in the data.
For each spectral point $i$, the derivative of Stokes $I$ with respect
to the wavelength was evaluated as

\begin{equation}
\left( \frac{{\rm d}I}{{\rm d}\lambda} \right)_{\lambda=\lambda_i}
= \frac{N_{i+1}-N_{i-1}}{\lambda_{i+1}-\lambda_{i-1}},
\label{eqn:four}  
\end{equation}

where $N_i$ is the photon count at wavelength $\lambda_i$.
As noise strongly influences the derivative, 
we interpolate the data after spectrum extraction with splines.
In our calculations we assumed a Land\'e factor $g_{\rm eff}$=1 for
hydrogen lines.
We are using 23 lines of \ion{He}{i}, \ion{He}{ii}, \ion{C}{iii},
\ion{C}{iv}, \ion{N}{ii}, \ion{N}{iii}, and \ion{O}{iii} in our analysis. 
The average Land\'e factor of these lines for measurements carried out with GRISM 600B is
$g_{\rm eff}$=1.07 and the average Land\'e factor of these lines for measurements carried out
with GRISM 600R is $g_{\rm eff}$=1.02.
More details of the observing technique are given by Bagnulo et al.\ (\cite{bagnulo:2002}) and 
Hubrig et al.\ (\cite{hubrig04a}, \cite{hubrig04b}).

\begin{figure}
\centering
\includegraphics[width=0.45\textwidth]{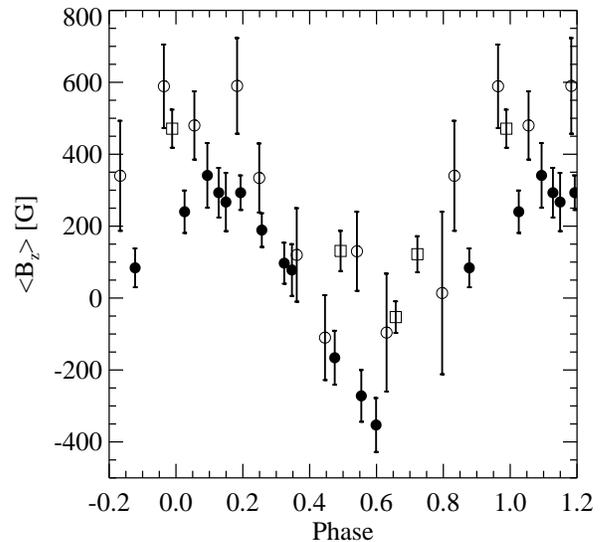}
\caption{
$\left<B_z\right>$ vs.\ the rotation phase for $\theta^1$\,Ori\,C.
Open circles: Observations by Wade et al.\ (\cite{wade06}) with MuSiCoS.
Open squares: Observations by Petit et al.\ (\cite{pet08}) with ESPaDOnS and Narval.
Filled circles: Our FORS\,1 measurements.
For one measurement of $\theta^1$\,Ori\,C presented by Petit et al.\ (\cite{pet08})  the phase 0.05 seems to be erroneous. 
Using the HJD of the observations presented in their Table~1, we calculate the phase 0.99.
}
\label{fig:theori_phasemap}
\end{figure}

Longitudinal magnetic fields were measured in two ways: using only the absorption hydrogen Balmer 
lines or using the whole spectrum including all available absorption lines of hydrogen,
\ion{He}{i}, \ion{He}{ii}, \ion{C}{iii}, \ion{C}{iv}, \ion{N}{ii}, \ion{N}{iii} and \ion{O}{iii}.
The lines that show evidence for emission were not used in the 
determination of the magnetic field strength (see Sec.\,3).
The feasibility of longitudinal magnetic field measurements in massive stars 
using FORS\,1 in spectropolarimetric mode was demonstrated by recent studies of early B-type stars
(e.g., Hubrig et al.\ \cite{hubrig:2006}, Hubrig et al.\ \cite{hubrig07}, 
Hubrig et al.\ \cite{hubrig08}).
In Fig.~\ref{fig:theori_phasemap} we demonstrate the excellent potential of FORS\,1 for measuring magnetic fields
in the star $\theta^1$\,Ori\,C, which was the first O-type star with a detected weak magnetic field varying with the rotation 
period of 15.4\,days. The open symbols represent previous magnetic field measurements 
by other authors. It is obvious that the FORS\,1 measurements are sufficiently accurate, showing a smooth 
sinusoidal curve in spite of the phase gap between 0.60 and 0.88. The values for the measured longitudinal 
magnetic field in different rotational phases are presented in Table~\ref{tab:theori_magfield}.

However, our observations determine a magnetic geometry different from the one deduced by
Wade et al.\ (\cite{wade06}). 
The maxima and minima of the measured longitudinal field as well as the phases of the field extrema appear 
to be different.
We are not aware of any systematic shift between FORS\,1 measurements 
and measurements with other spectropolarimeters. We periodically observe well-studied magnetic stars with
known variation curves and usually our measurements are 
in good agreement with those obtained  with other instruments.
On the other hand, the reason for such a shift is understandable since Wade et al.\ (\cite{wade06}) used for their 
measurements just three metal lines, \ion{O}{iii} 5592, \ion{C}{iv} 5801,
and \ion{C}{iv} 5811.
The profiles of these lines exhibit clear variations which could be 
signatures of an uneven distribution of these elements over the stellar surface
(see e.g.\ Reiners et al.\ \cite{reiners00}).
Such an uneven element distribution of metal lines will 
affect the line-of-sight component of the magnetic field integrated over the stellar surface. 
A different set of lines is of special
relevance for magnetic field measurements only in the cases where few lines are used.
If the distribution 
of spots of different elements on the stellar surface is related to the magnetic field geometry (as 
is usually found in classical magnetic Ap and Bp stars where certain elements are concentrated on magnetic poles 
and other elements along the stellar magnetic equator), magnetic 
field measurements using the lines of different elements will produce different magnetic field strengths,
depending on the location of the elemental spots on the stellar surface.
With FORS\,1 we use for the measurements all absorption lines belonging to various chemical elements together,
and in this way might sample the magnetic field more uniformly over the observed hemisphere.

Without further detailed 
high-resolution studies of polarized line profiles of different elements it is currently not 
obvious which set of measurements lies closer to the true longitudinal magnetic field of  $\theta^1$\,Ori\,C.
Assuming an inclination of the rotation axis to the line-of-sight of $i$=45$^\circ$ (Wade et al.\ \cite{wade06}),
our modeling of the longitudinal field variation constrains the dipole 
magnetic field geometry of  $\theta^1$\,Ori\,C to $B_d\approx1100$\,G and $\beta$ close to $90^\circ$,
where $B_d$ is the dipole intensity and $\beta$ is the obliquity angle.

\section{Results}
\label{sect:results}

\begin{figure*}
\centering
\includegraphics[width=\linewidth,height=0.666\linewidth]{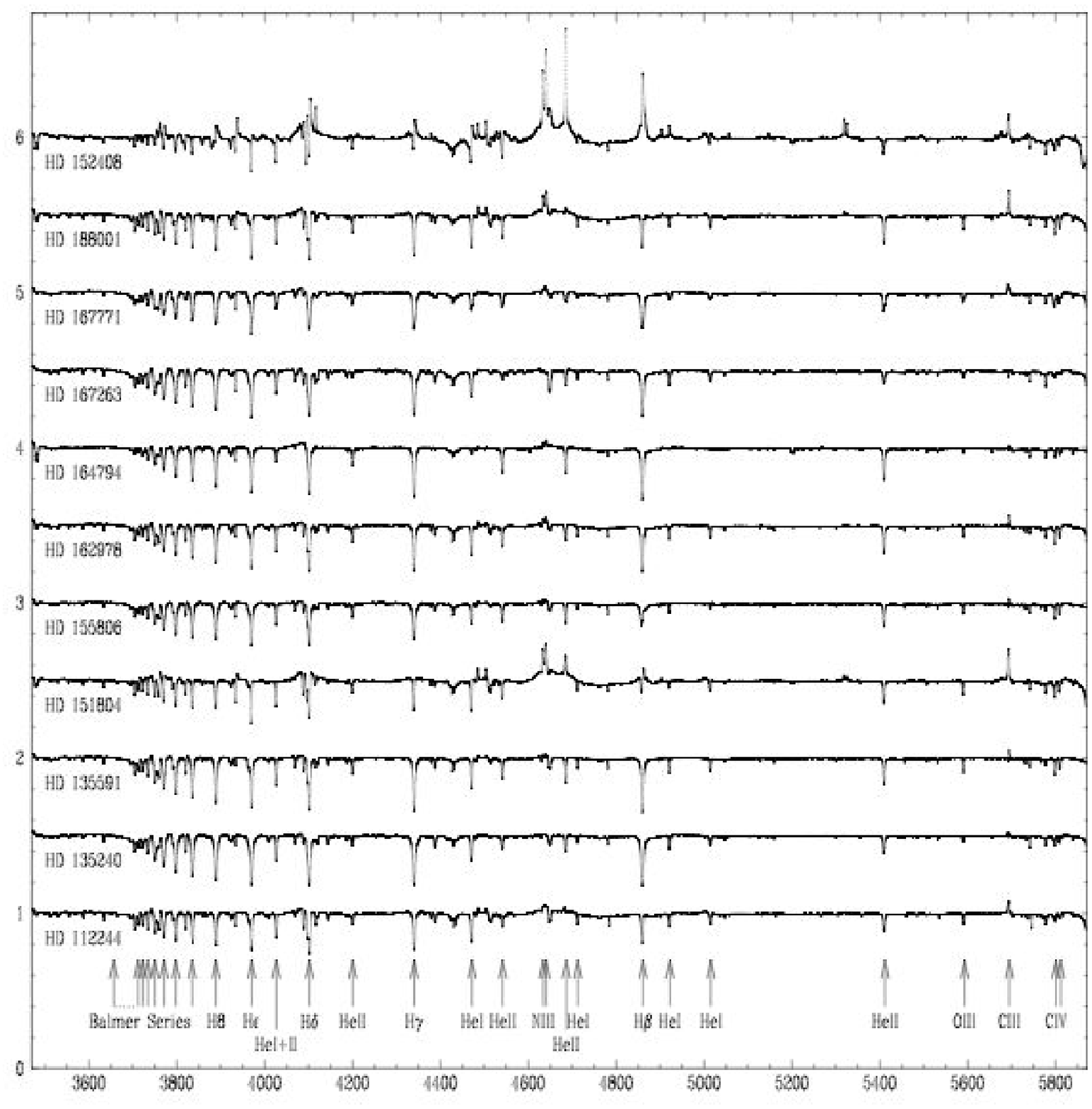}
\caption{
Normalised FORS\,1 Stokes~$I$ spectra of O-type stars observed in 2005.
Well known spectral lines have been indicated by the arrows, all Balmer lines from the Balmer jump to H$\beta$ are visible.
The spectra were offset from 1 by multiples of 0.5 for clarity.
}
\label{fig:spectra}
\end{figure*}

\begin{table}
\centering
\caption{
Magnetic field measurements of $\theta^1$\,Ori\,C with FORS\,1.
Phases are calculated according to the ephemeris of 
Stahl et al.\ (\cite{stahl:1996}), ${\rm JD} = 2\,448\,833.0 + 15.422~{\rm E}$.
All quoted errors are 1$\sigma$ uncertainties.
}
\label{tab:theori_magfield}
\begin{tabular}{lcr@{$\pm$}l}
\hline
\multicolumn{1}{c}{MJD} &
\multicolumn{1}{c}{Phase} &
\multicolumn{2}{c}{$\left<B_z\right>$}\\
 &
 &
\multicolumn{2}{c}{[G]} \\
\hline
\hline
54107.221 & 0.0257 & 240 & 59 \\
54108.272 & 0.0939 & 341 & 90 \\
54109.127 & 0.1494 & 267 & 81 \\
54112.174 & 0.3469 & 78 & 72 \\
54114.149 & 0.4750 & $-$166 & 75 \\
54116.057 & 0.5987 & $-$353 & 75  \\
54155.062 & 0.1279 & 293 & 69 \\
54156.072 & 0.1933 & 293 & 48 \\
54157.051 & 0.2569 & 189 & 47 \\
54158.086 & 0.3239 & 97 & 57 \\
54177.064 & 0.5545 & $-$272 & 72 \\
54182.048 & 0.8777 & 84 & 54 \\
\hline
\end{tabular}
\end{table}

\begin{table}
\caption{
Longitudinal magnetic fields measured with FORS\,1 in 13 O-type stars.
All quoted errors are 1$\sigma$ uncertainties.
}
\label{tab:fields}
\centering
\begin{tabular}{rcr @{$\pm$} lr @{$\pm$} lc}
\hline
\hline
\multicolumn{1}{c}{HD} &
\multicolumn{1}{c}{MJD} &
\multicolumn{2}{c}{$\left< B_z\right>_{\rm all}$} &
\multicolumn{2}{c}{$\left< B_z\right>_{\rm hydr}$} &
\multicolumn{1}{c}{Comment} \\
 &
 &
\multicolumn{2}{c}{[G]} &
\multicolumn{2}{c}{[G]} &
 \\
\hline
36879  & 54345.389       & {\bf 180}    & {\bf 52} & 109          & 74       & ND \\
112244 & 53455.193       & 34           & 55       & 15           & 62       & \\
       & 53475.177       & 41           & 43       & 1            & 60       & \\
       & 53483.104       & 9            & 78       & $-$4         & 79       & \\
135240 & 53475.246       & 65           & 83      & 86           & 111      & \\
       & 53487.263       & $-$37        & 62       & $-$12        & 72       & \\
       & 53553.103       & $-$65        & 63       & $-$45        & 78       & \\
135591 & 53487.243       & $-$118       & 57       & $-$142       & 62       & \\
       & 53553.081       & 110          & 54       & 116          & 61       & \\
       & 53571.081       & $-$8         & 62       & $-$20        & 71       & \\
148937 & 54550.416       & {\bf $-$276} & {\bf 88} & $-$145       & 104      & ND \\
151804 & 53476.369       & $-$151       & 90       & $-$87        & 96       & \\
       & 53571.025       & 68           & 65       & 91           & 73       & \\
       & 53596.061       & 82           & 46       & 66           & 48       & \\
152408 & 53556.216       & {\bf $-$89}  & {\bf 29} & $-$112       & 57       & ND \\
       & 53571.104       & $-$91        & 46       & $-$93        & 75       & \\
       & 53596.081       & 46           & 34       & 32           & 60       & \\
155806 & 53476.401       & $-$80        & 132      & $-$216       & 141      & \\
       & {\it 53532.283} & {\it $-$115} & {\it 37} & {\it $-$119} & {\it 50} & PD\\
       & 53532.306       & $-$29        & 44       & $-$35        & 70       & \\
       & 53556.235       & $-$184       & 88       & $-$160       & 93       & \\
       & 54549.403       & 93           & 68       & 54           & 88       & \\
162978 & 53556.260       & $-$50        & 49       & $-$56        & 86       & \\
       & 53595.116       & 91           & 81       & 73           & 84       & \\
       & 53604.144       & 80           & 83       & 60           & 89       & \\
164794 & 53520.357       & $-$114       & 66       & $-$111       & 75       & \\
       & 53594.119       & {\bf 211}    & {\bf 57} & 147          & 72       & ND \\
       & 53595.096       & $-$165       & 75       & $-$139       & 77       & \\
167263 & 53594.142       & $-$24        & 91       & $-$54        & 96       & \\
       & 53595.015       & $-$19        & 41       & $-$29        & 49       & \\
       & 53596.112       & 29           & 53       & 37           & 61       & \\
167771 & 53520.377       & 5            & 79       & 11           & 85       & \\
       & 53594.241       & 92           & 46       & 78           & 73       & \\
       & 53595.066       & $-$31        & 54       & $-$16        & 88       & \\
188001 & 53520.434       & 117          & 65       & 100          & 65       & \\
       & 53594.208       & $-$35        & 50       & $-$53        & 55       & \\
       & 53595.149       & $-$35        & 36       & $-$32        & 57       & \\
       & 53597.149       & $-$95        & 48       & $-$163       & 70       & \\
\hline
\end{tabular}
\end{table}

Because of the strong dependence of the longitudinal magnetic field on the rotational 
aspect, its usefulness to characterise actual magnetic field strength distributions
depends on the sampling of the various rotation phases,
hence various aspects of the magnetic field. 
All targets were observed on three or four different nights. As mentioned before, the 
exceptions are the stars HD\,36879 and HD\,148937, which we have been able to observe only once.
Apart from HD\,148937 which has a rotation period of seven days (Naz\'e et al.\ \cite{naze08}), 
no exact rotation periods are known for the other stars in our sample, 
and certainly, it is not possible to characterise the magnetic field topology
with only a few measurements. We would like to emphasize on the other hand that since 
the existence of magnetic fields
in O stars has been suspected for a long time for many compelling reasons, including recent 
theoretical developments, already a mere discovery of such fields is of great importance, and their subsequent 
detailed studies will help to explain a wide range of well documented enigmatic phenomena in massive stars. 

Normalised FORS\,1 Stokes $I$ spectra of all targets observed in 2005 together with the identification of the strongest spectral lines are 
presented in Fig.~\ref{fig:spectra}.
Compared to lower mass stars less lines were available for the measurement of the magnetic field and the metal lines 
are not strong.
Even the strongest hydrogen lines have a maximum depth of only about 40\% below the continuum,
as they are intrinsically weaker than in the B and A type stars.
P~Cygni profiles and pure emission lines are visible in all stars apart from HD\,167263 which shows 
a weak emission only in the C~III line at 5696\,\AA{}
(only marginally visible
at the resolution provided in this paper).
This C~III line is in emission in the spectra of all studied stars.
As we emphasised in the previous section, the lines that show evidence for emission were not used in the 
determination of the magnetic field strength.
Most of  these lines are wind-formed and may have very different polarisation signatures.

The results of our magnetic field measurements are presented in Table~\ref{tab:fields}.
In the first two columns we give the HD numbers of the targets and the modified Julian dates of the middle of the exposures. 
The measured mean longitudinal magnetic field $\left<B_{\mathrm z}\right>$ using all absorption lines 
is presented in column~3.
%
The measured mean longitudinal magnetic field $\left<B_{\mathrm z}\right>$ using all hydrogen lines in 
absorption is listed in column~4.
All quoted errors are 1$\sigma$ uncertainties.
In column~5 we identify new detections by ND.
We note that all claimed detections have a significance of at
least 3$\sigma$, determined from the formal uncertainties we derive. These measurements are indicated in bold face.

Four stars of our sample, HD\,36879, HD\,148937, HD\,152408, and HD\,164794, show evidence for the presence of a weak magnetic 
field in the measurements using all spectral absorption lines. The uncertainties of the 
mean longitudinal field determination is obtained from the formal uncertainty of the linear regression
of $V/I$
versus the quantity
$-\frac{g_{\rm eff}e}{4\pi{}m_ec^2} \lambda^2 \frac{1}{I} \frac{{\mathrm d}I}{{\mathrm d}\lambda} \left<B_z\right> + V_0/I_0$.
For measurements obtained from Balmer lines only, the mean uncertainty ranging from 49 to 141\,G is generally larger than 
for measurements  using all absorption lines where the uncertainty can be as small as 29\,G. 
These results are easily understandable since the robustness and accuracy of the 
spectropolarimetric observations increase with the number of spectral lines used for the measurements.
In previous studies it was shown that
the uncertainty of the FORS\,1 measurements can be as low as 13\,G in late A spectral types with numerous strong 
absorption lines (e.g.\ Kurtz et al.\ \cite{kurtz08}).

The stars HD\,36879 and HD\,148937 were observed only once and their distinct magnetic field is detected at the 3.5 and 
3.1$\sigma$ level, respectively.
Since the rotation period of HD\,148937 is known (Naz\'e et al.\ \cite{naze08}), the  $v\sin i$ value 
is relatively low and a comparatively large magnetic field is detected,
this star should clearly be of highest priority 
for future spectropolarimetric observations over the rotation period to study its magnetic field 
topology. 
In Table~\ref{tab:fields} the second observation of HD\,152408 reveals a magnetic field at 
almost 2$\sigma$ level, and the third observation
of HD\,164794 shows a magnetic field at 2.2$\sigma$ level. We should note that the marginal detections 
of the magnetic field 
for HD\,152408 and HD\,164794 on these observing nights can naturally be explained by the 
strong dependence of the longitudinal magnetic field on the rotational aspect.
In case of $\theta^1$\,Ori\,C, out of the ten magnetic field measurements presented by Wade et al.\ (\cite{wade06}), 
only four measurements have values larger than 3$\sigma$, and among the four measurements of Petit et al.\ (\cite{pet08}) 
only one measurement is at a high significance level. Among our 12 observations of $\theta^1$\,Ori\,C, four 
measurements could be considered as marginal detections. However, all measurements plotted over the rotational 
phase of 15.4\,d can be well--presented by a sine fit characterising a dipole magnetic field 
of a certain pole strength and inclination to the rotational axis. 

Interestingly, most stars in our sample which have been observed on different nights show a change of polarity.
The star HD\,188001, which was
observed on four different nights, shows one 2.3$\sigma$ detection obtained from Balmer lines and 
2$\sigma$ detection using all absorption lines.
The magnetic field of the star HD\,135591 was observed at 2.1$\sigma$ and 2$\sigma$ levels on two different
nights using all absorption lines.
Both stars seem to be good candidates for future magnetic field measurements.
The star HD\,155806 was already observed once with FORS\,1 using GRISM 1200g in the framework of the ESO service 
program 075.D--0507 by Hubrig et al.\ (\cite {hubrig07}), who reported the presence of a 
weak mean longitudinal field, $\langle$$B_z$$\rangle$\,=\,$-$115$\pm$37\,G.
For convenience the previously published measurement is presented in the same table in italics and marked as
PD (previous detection). 
Four new  measurements with the GRISM 600B show a polarity change, but all of them are marginal detections.
The first measurement has in addition very low accuracy  due to bad weather 
conditions during service observations. One of the measurements reveals a longitudinal magnetic 
field at the 2.1$\sigma$ level: $\langle$$B_z$$\rangle$\,=\,$-$184$\pm$88\,G.
Although we do not detect magnetic fields for other
O-type stars at a 3$\sigma$ level, it is still possible that some of them host magnetic fields, but that these 
fields remain undetected due to rather high measurement uncertainties.
We note that further observations with an improved 
accuracy are clearly necessary in order to put an upper limit on the strength of their longitudinal 
magnetic fields.

Appendix~\ref{sect:app} provides a brief overview of the previous knowledge
of the stars with a magnetic field detection at 
3$\sigma$ level. A few notes are also given on HD\,155806 for which the presence of a magnetic field 
was recently reported by Hubrig et al.\ (\cite {hubrig07}).

\section{Discussion}
\label{sect:discussion}

Stellar magnetic fields have been discovered across a large range of spectral types 
(see Charbonneau \& MacGregor \cite{charbonneau:2001}).
In late type stars, dynamos active in the convective layers are believed to be the origin of the observed magnetic fields.
In earlier type stars, which have radiative envelopes, large scale magnetic fields of the order of a 
kilogauss have been discovered in Ap/Bp stars,
but the exact origin of these fields is not yet known
(Charbonneau \& MacGregor \cite{charbonneau:2001}, Braithwaite \& Nordlund \cite{braithwaite:2006}). 
About 10\% of main-sequence A and B stars are slowly 
rotating, chemically peculiar, magnetic Ap and Bp stars and among their descendants, 
white dwarfs, 10\% have high magnetic fields.
The magnetic fields in magnetic white dwarfs could be 
fossil remnants from the main-sequence phase, consistent with magnetic flux conservation
(Ferrario \& Wickramasinghe \cite{ferrario05}).
If we assume that massive stars behave like Ap and Bp stars, then for a magnetic field 
detection probability of 10\%, an O stars sample should consist of a larger number of 
unbiased targets, including  stars in clusters and the Galactic field
at different ages, in different Galactic metallicity zones,
and with different rotational velocities and surface composition.
As we mention in Sec.\,2, our sample is biased in the sense that it is restricted to O-type stars 
exhibiting wind variability, anomalous X-ray behaviour and brightness variations.

Still, this is the first time that magnetic field strengths were determined for such
a large sample of stars, with an accuracy comparable to the errors obtained for the three 
previously known magnetic O-type stars, $\theta^1$\,Ori\,C, HD\,155806, and HD\,191612. For
the magnetic Of?p star HD\,191612, Donati et al.\ (\cite{donati:2006}) measured a magnetic field of $\langle B_\mathrm
{z}\rangle=-220\pm38$\,G, averaging a total of 52 exposures obtained over 4 different nights. This is similar 
to our typical errors of a few tens of G.
We have found four new magnetic O-type stars which have different spectral types, luminosity 
classes, and behaviour in various observational domains (see Appendix~\ref{sect:app}). It opens the question 
whether O-type stars are magnetic in different evolutionary states.
The study of the evolutionary state of one of the  Galactic Of?p stars, HD\,191612, indicates 
that it is significantly 
evolved with an $\sim$O8 giant-like classification (Howarth et al.\ \cite{how07}).
The youth of the best studied magnetic O-type star $\theta^1$\,Ori\,C and the older age 
of the Of?p star HD\,191612 suggest that the presence of magnetic fields in O-type stars is 
not related to their evolutionary state. On the other hand, $\theta^1$\,Ori\,C seems 
to possess a somewhat stronger magnetic field in comparison to that of HD\,191612, indicating 
that the magnetic field could be a fossil remnant. Considering different size of radii 
due to the different evolutionary state and assuming conservation of the magnetic surface flux 
the magnetic field strength is expected to decrease by a factor proportional to the square
of the ratio of their radii.
Some support for the fossil magnetic field origin arises also from the recent 
comparative study of magnetic fields of early B-type stars at different ages by 
Briquet et al.\ (\cite{bri07}). This work revealed that the strongest 
magnetic fields appear in the youngest Bp stars, compared to weaker magnetic fields in stars 
at advanced ages.  

It is notable, that the current analyses of Naz\'e et al.\ (\cite{naze08}) and Naz\'e et al.\ (in preparation)
suggest slower rotation than usually observed in O-type stars and nitrogen enhancement in 
both other Of?p stars, HD\,108 and HD\,148937.
Their multiwavelength studies indicate rotation periods of $\sim$55 yr for HD\,108 
and 7\,d for HD\,148937.
Also, recent NLTE abundance analyses of early B-type stars by  Morel et al.\ (\cite{mor08}) and 
Hunter et al.\ (\cite{hunt08})
confirm that slow rotators often have peculiar chemical enrichments.
It is remarkable that the observational data collected by
Morel et al.\ (\cite{mor08})
strongly point to a higher incidence of magnetic fields
in stars with nitrogen excess and boron depletion.
Clearly, future studies are necessary to determine the efficiency of various indirect 
indicators for the presence of magnetic fields in the atmospheres of hot stars.

Yet, although it was possible to recognise a few hot magnetic stars as peculiar 
from their spectral morphology prior to their field detection (Walborn \cite{walborn06}), 
the presence of a magnetic field can also be expected in stars of other classification 
categories. Our measurements of 13 O-type stars indicate that magnetic fields are possibly present 
in stars with very different behaviour in visual, X-ray and radio domains.
Subsequent magnetic field measurements will test and constrain the conditions controlling the 
presence of magnetic fields in hot stars, and the implications of such fields on their 
mass-loss rate and evolution.

Since no longitudinal magnetic fields stronger than 300\,G were detected in our study
(apart from $\theta^1$\,Ori\,C) we suggest that
large scale, dipole like, magnetic fields with polar field strengths
larger than 1\,kG are not widespread among O type stars.
Stars more massive than about 9~M$_\odot$ end up as neutron stars or as black holes.  
A significant fraction of newborn neutron stars are strongly magnetized, with typical 
fields of $\sim10^{12}$\,G, and fields of up to $\sim10^{15}$\,G in the magnetars. 
Simple conservation of magnetic flux would imply field strengths of at least 
(5\,R$_\odot$/10\ km$)^{-2}\times10^{12}$\,G $\simeq 10^1$\,G as a minimum for their progenitors. 
This is similar to the minimum field strength required to explain the wind variability 
observed in the UV (several $10^1$ G), as can be concluded from numerical simulations 
of wind behaviour in early type stars (ud-Doula \& Owocki \cite{asif:2002}). Our measurements have a typical accuracy of a
few tens of G and it is quite possible that weak magnetic fields are present in the atmospheres of the 
other stars of our sample, but remain undetected as long as the measurement uncertainties are 
not significantly improved. 

As we mention in Sec.\,2, a new mosaic detector
with blue optimised E2V chips was implemented in FORS\,1 in 2007.
To achieve the highest possible signal-to-noise (S/N) ratio -- as is 
required for accurate measurements of stellar magnetic fields --
the (200kHz, low, 1$\times$1) readout mode can be used, which makes it possible to 
achieve a S/N  ratio of more than 1000 with only one single sub-exposure. Our recent tests 
show that using a sequence of 8--10 sub-exposures we can obtain 
much better accuracies down to 10--20\,G.
The typical uncertainties presented in Table 2 and 3 at the time of our observations before 
this CCD upgrade have been of the order of 40 to 70\,G.

In the absence of further direct magnetic measurements, it is not clear yet whether 
more complex, smaller scale fields play a role in the atmospheres of O-type stars.
In the case of a more complex magnetic field topology the longitudinal magnetic field  
integrated over the visible stellar surface 
will be smaller (or will even cancel) and not be
easily detected with the low-resolution FORS\,1 measurements, which allow to detect
only magnetic fields which possess a large-scale organization.
However, high resolution spectropolarimeters should be able to detect such complex field configurations
using high signal-to-noise observations of the Zeeman effect in metal lines (e.g.
Donati et al.\ \cite{donati:2006b}).

\appendix

\section{Brief notes on stars with a possible evidence for the presence of a magnetic field}
\label{sect:app}

\begin{figure*}
\centering
\includegraphics[height=0.45\textwidth,angle=270]{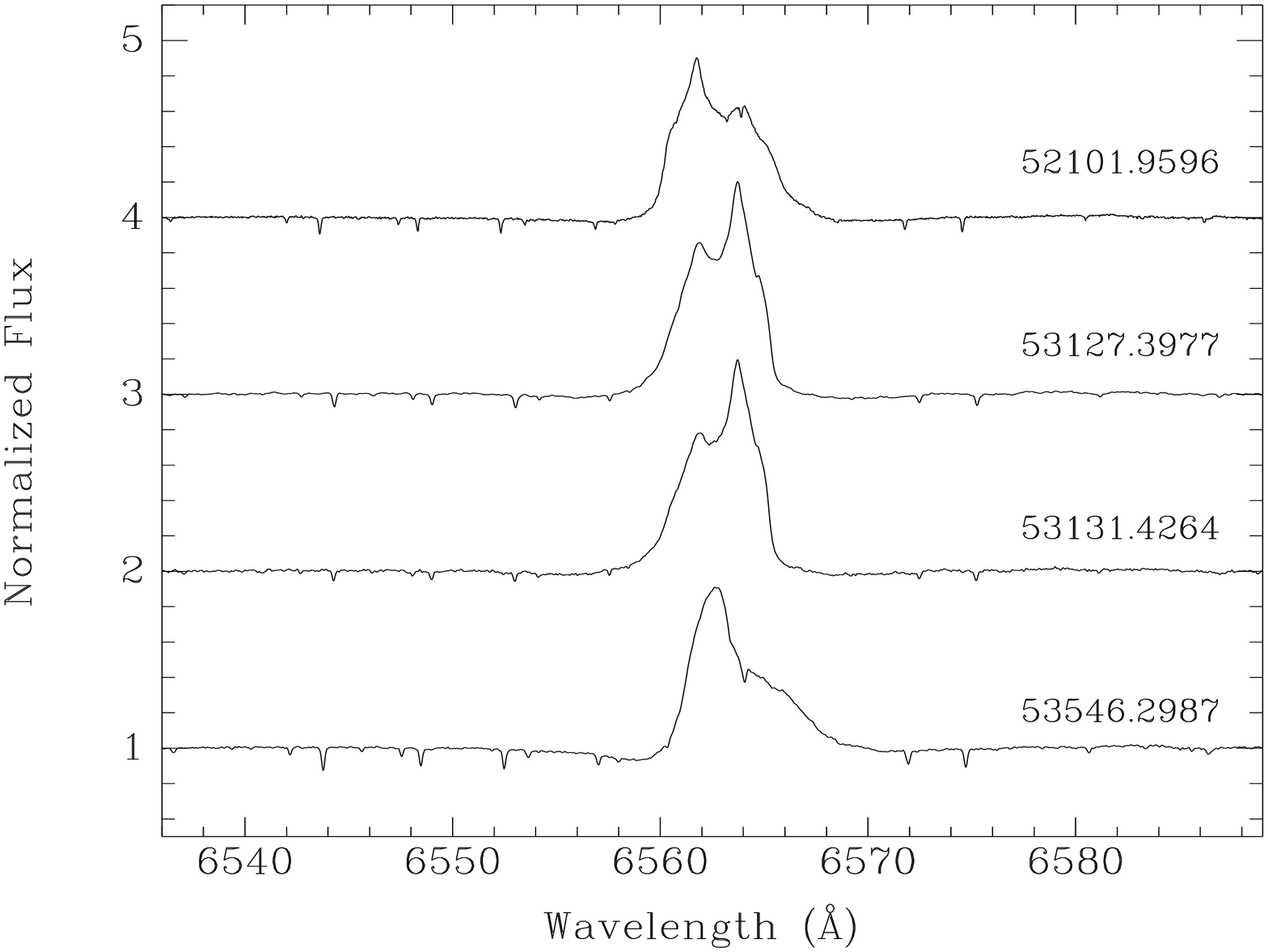}
\includegraphics[height=0.45\textwidth,angle=270]{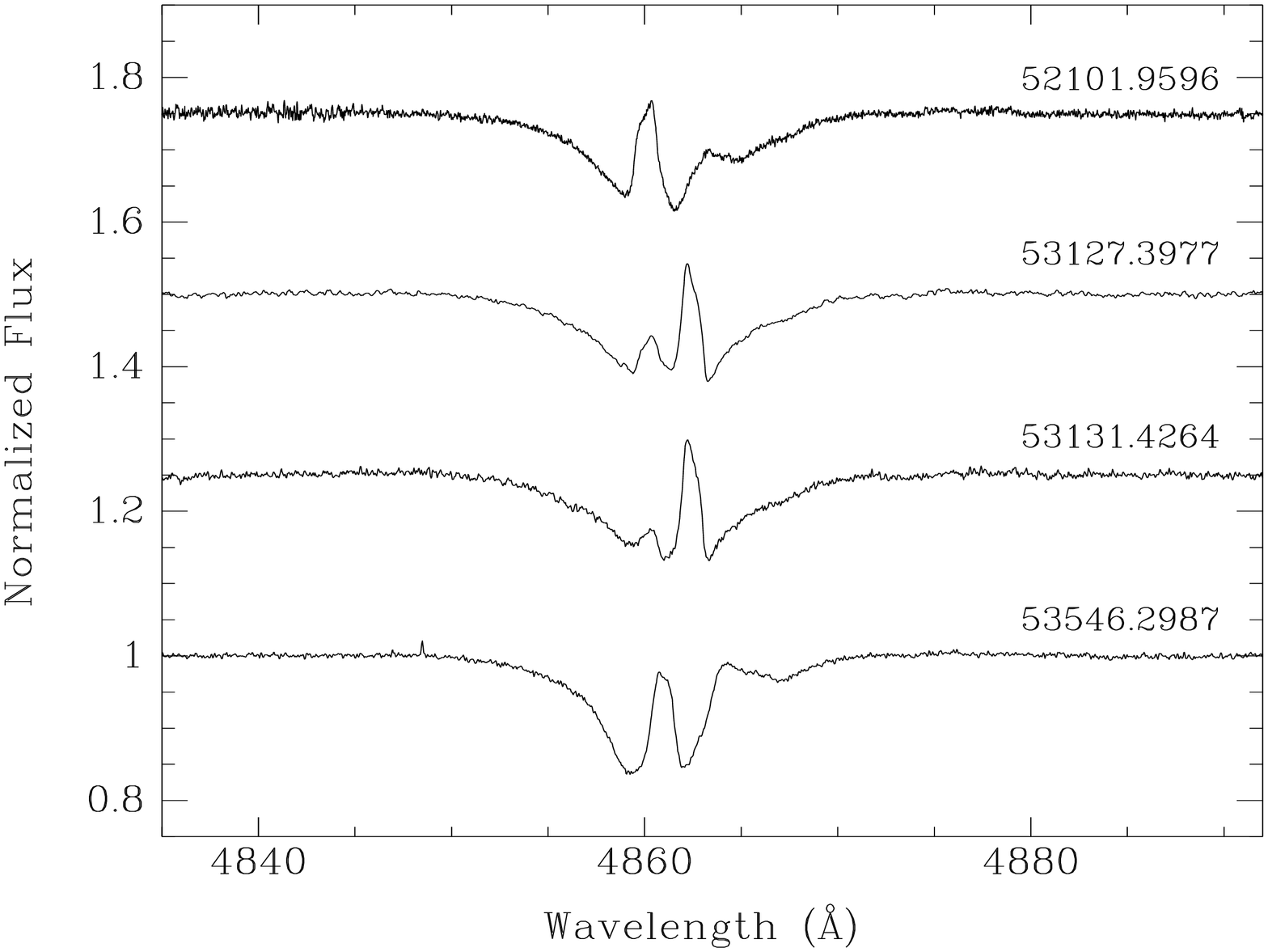}
\caption{
Spectral profile variability of the Balmer H$\alpha$ and H$\beta$ lines in the FEROS and UVES spectra of HD\,155806.
The spectra are labeled with their modified Julian dates.
}
\label{fig:Halphabeta}
\end{figure*}

\begin{figure*}
\centering
\includegraphics[height=0.45\textwidth,angle=270]{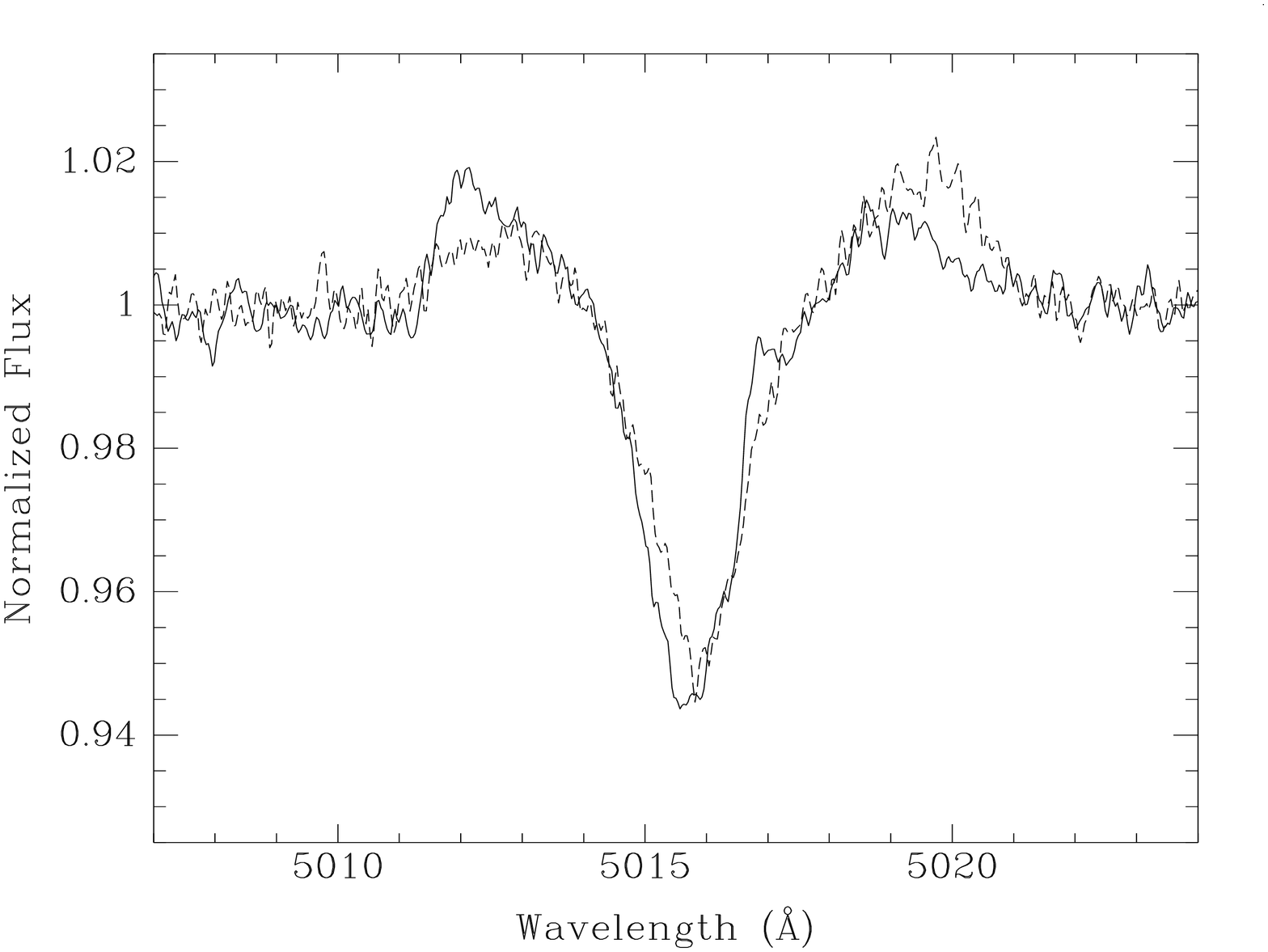}
\includegraphics[height=0.45\textwidth,angle=270]{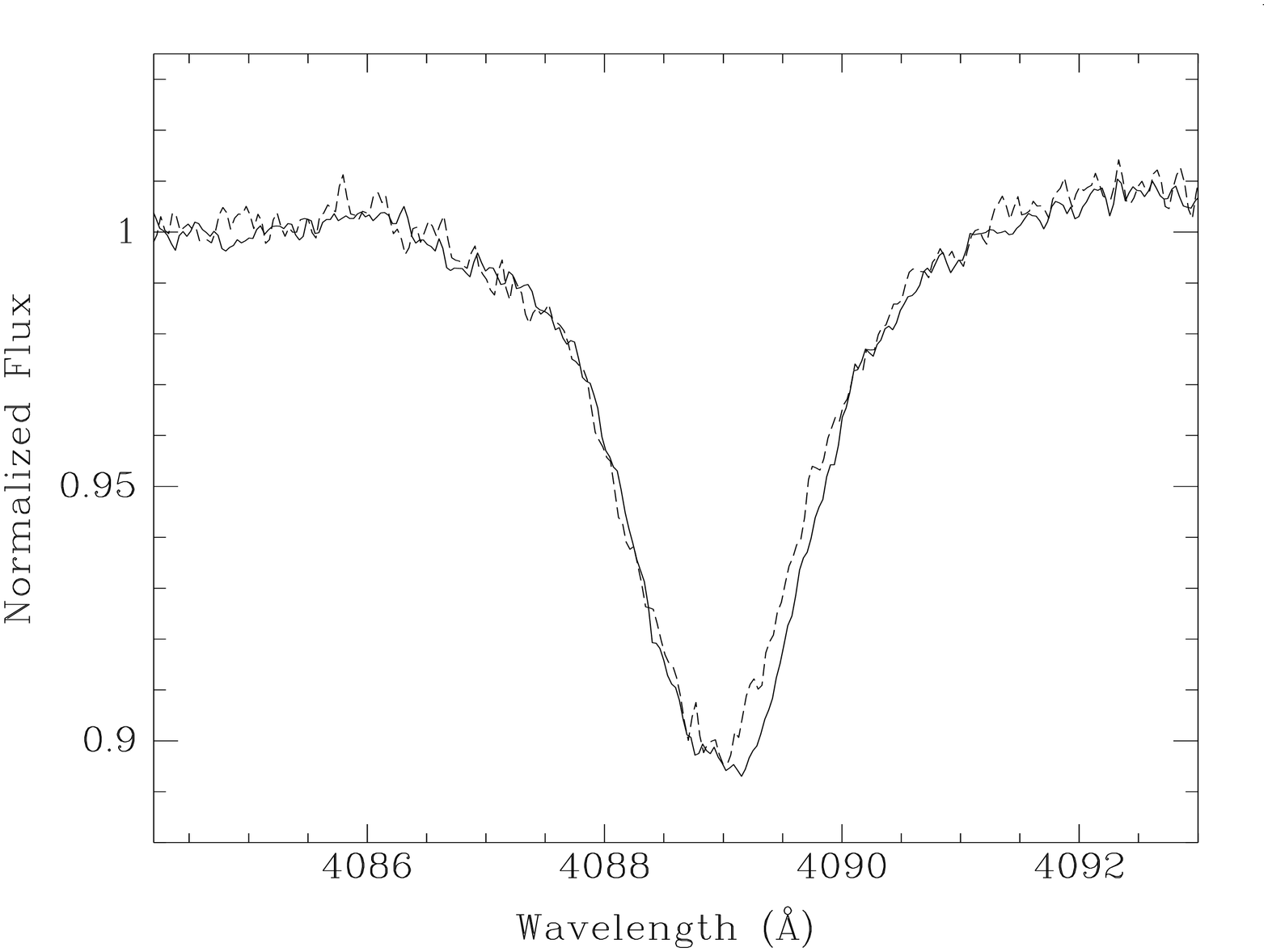}
\caption{
Small--scale variations of the \ion{He}{i} 5016\,\AA{} and \ion{Si}{iv} 4089\,\AA{} lines in the spectra of 
HD\,155806 obtained on the modified Julian dates 53127.40 (solid line) and 53546.30 (dashed line). 
}
\label{fig:HeSi}
\end{figure*}

\subsection{HD\,36879}
This star has a classification O7~V(n) according to the Galactic O Star Catalogue 
(Ma{\'{\i}}z-Apell{\'a}niz et al.\ \cite{maiz:2004}). It exhibits peculiar narrow
emission features in the \ion{Si}{iv} line profiles $\lambda$$\lambda$1394, 1403\,\AA{} detected by Walborn \& Panek (\cite{walb84a})
and Walborn \& Panek (\cite{walb84b}). The authors noted that these emission features are strongly variable on the IUE 
spectra obtained four days apart. Otherwise, this star is only marginally studied, mainly due to its rather fast 
rotation.

\subsection{HD\,148937}
Only three Galactic Of?p stars are presently known: HD\,108, HD\,148937, and HD\,191612.
No attempt has been made yet to measure the magnetic field in HD\,108.
The search for a magnetic field in HD\,191612 was motivated by very unusual,
large, periodic spectral variations found previously by Walborn et al.\ (\cite{walb03})
and resulted in $\langle$$B_z$$\rangle$\,=\,$-$220$\pm$38\,G,
measured with ESPaDOnS (Donati et al.\ \cite{donati:2006}).
An extensive multiwavelength study of HD\,148937 was recently carried out by Naz\'e et al.\ (\cite{naze08}) who
detect small-scale variations of \ion{He}{ii} 4686 and Balmer lines with a period of 7\,days and an overabundance 
of nitrogen by a factor of 4 compared to the Sun.

\subsection{HD\,152408}
This star is a member of NGC\,6231 and was classified as O8 Iafpe or WN9ha by Walborn \& Fitzpatrick (\cite{walb00}).
Observational studies of wind and photospheric variability have been performed by Colley (\cite{colley02})
and Prinja et al.\ (\cite{prinja01}).
The interesting fact is that Prinja et al.\ (\cite{prinja01}) discovered that the line profile behaviour 
was clearly not erratic, but instead organised into sequential localised episodes of enhanced and/or 
reduced flux, which migrated in velocity as a function of time. They also demonstrated that systematic variability 
is present in absorption lines formed in the photospheric layers of this star and suggest that the presence of a 
magnetic field in particular may provide significant variation in the mass-flux and thus also account for the 
fluctuations discovered in the central regions of the H$\alpha $ emission line. 
A medium-resolution spectropolarimetric study of the H$\alpha$ emission line by Harries et al.\ (\cite{har02})
showed that the continuum polarisation agrees well with the local interstellar polarisation pattern.
However there appears to be 
slight evidence of a position angle rotation in combination with a magnitude enhancement across H$\alpha$.

\subsection{HD\,155806}
The star HD\,155806 has been classified as O7.5~V[n]e by Walborn (\cite{walb73}), but was reclassified by 
Negueruela et al.\ (\cite{neg04}) as  O7.5~IIIe
based on the strength of its metallic features.
A strong variability of Balmer lines and small-scale variations of \ion{Si}{iv} and \ion{He}{i} lines
have been detected in FEROS and UVES spectra retrieved from the ESO archive 
(Programs 073.C--0337, 073.D--0609, 075.D--0061, and 266.D--5655).
Spectral profile variations of H$\alpha$ and H$\beta$ lines are presented in Fig.~\ref{fig:Halphabeta} and 
those of the \ion{He}{i} 5016\,\AA{} and  \ion{Si}{iv} 4089\,\AA{} lines in  Fig.~\ref{fig:HeSi}.

\subsection{HD\,164794}
This star has a classification O4~V((f)) according to the Galactic O Star Catalogue (Ma{\'{\i}}z-Apell{\'a}niz et al.\ \cite{maiz:2004}),
exhibiting weak \ion{N}{iii} emission and strong \ion{He}{ii} $\lambda$4686 absorption. 
It is a well-known non-thermal radio emitter and according to van Loo et al.\ (\cite{van06}) the most 
likely mechanism is synchrotron emission from colliding winds, implying that all O stars 
with non-thermal radio emission should be members of binary or multiple systems.
Hints of a wind-wind interaction were indeed detected in the X-ray 
domain (Rauw et al.\ \cite{rauw02}).
Currently, a long-term study of its binary 
nature and spectrum variability is undertaken by our Belgian colleagues 
(see preliminary results in Rauw et al.\ \cite{rauw05}).

{
\acknowledgements
This research has made use of VALD, ADS, and SIMBAD.
We would like to thank R.~Mokiem for help on theoretical spectra for O stars,
M. Briquet for participating in the observing run 079.D-0241, T. Szeifert for providing the
FORS\,1 spectra extraction routines, and Y.~Naz\'e, N.R.~Walborn and the anonymous referee for valuable comments.
}


\end{document}